\documentclass[preprint,aps,amsmath]{revtex4}

\usepackage{graphicx}
\usepackage{amsmath}
\usepackage{bm}

\newcommand\bnabla{\boldsymbol \nabla}

\begin{document}
\preprint{}

\title{Modelling of turbulent impurity transport in fusion edge plasmas\\
using  measured and calculated ionization cross sections}

\author{Alexander Kendl}

\affiliation{Institute for Ion Physics and Applied Physics, University of
  Innsbruck,\\ Association Euratom-\"OAW, Technikerstr. 25, 6020 Innsbruck, Austria\vspace{1.5cm}}

\begin{abstract}
\vspace{0.5cm}
Turbulent transport of trace impurities impurities in the edge and
scrape-off-layer of tokamak fusion plasmas is modelled by three dimensional
electromagnetic gyrofluid computations including evolution of plasma profile gradients. 
The source function of impurity ions is dynamically computed from pre-determined
measured and calculated electron impact ionization cross section data. 
The simulations describe the generation and further passive turbulent $E
\times B$ advection of the impurities by intermittent fluctuations and
coherent filamentary structures (blobs) across the scrape-off-layer.

\vspace{8.5cm}

\begin{center}
\noindent {\sl This is a preprint version (with reduced figure quality) of publication}\\ 
International Journal of Mass Spectrometry {\bf 365/366}, 106-113 (2014)\\
{\sl in an honorary issue dedicated to the 70th birthday of Tilmann M\"ark.}
\end{center}
\end{abstract}

\maketitle

\section{Introduction}
%\label{}

Magnetically confined plasmas for fusion research would ideally be composed of
electrons and only the main hydrogen ion species, which usually are either simply
protons or deuterium ions for most present experiments, or consist of a
deuterium and tritium ion fuel mixture and helium ions as their fusion product
in burning plasmas. 
The electron-hydrogen plasma unavoidably gets into contact with wall materials,
both (preferentially) in designated strike areas in the divertor region, and 
also (strongly undesired) on the first wall surrounding the bulk plasmas.

Plasma-wall interaction processes in tokamak fusion experiments
\cite{janeschitz01,federici03,roth09,clark05} can generate 
various neutral and ionized atoms and molecules, which may penetrate into and
profoundly disturb the main plasma, and also can lead to unfavourable
co-deposition of materials and composites (e.g. of tritiated hydrocarbons) in
other areas of the vessel.  
Detailed knowledge of atomic and molecular interactions in the edge of tokamaks
and of the transport of impurities is therefore of considerable
interest for understanding and modelling of fusion plasmas.

The major cross-field transport mechanism of particles and energy in magnetized
fusion plasmas is turbulent fluid-like convection by wave-like fluctuating
electric 
fields ${\bf E}({\bf x},t)$ acting on the plasma through the ${\bf E}
\times {\bf B}$ drift velocity in a background magnetic field  
\cite{tang78,hugill93,liewer85,wootton90,dimits00,scott03,garbet04,tynan09}.
The relevant drift wave turbulence micro-scales in tokamak edge plasmas are in the
order of sub-mm spatial vortex structures in the MHz frequency range, but are
further closely coupled to macro-scale zonal flow structures \cite{diamond05}
and meso-scale instabilities like edge-localized modes \cite{kendl10} and
magnetic islands.  

Once impurities, which are born at the outer edge of the plasma region, are
ionized by electron impact or other processes, they are also subjected to
the turbulent ${\bf E} \times {\bf B}$ advection and may as a consequence be
efficiently transported further across the field and radially inwards.

In this work we study the passive transport of fusion relevant trace impurity
ions by means of multi-species edge turbulence computations, which include
dynamical electron-impact ionization in fluctuating filamentary plasma
structures as a source for impurity ions in the scrape-off-layer region of a tokamak.

\section{Gyrofluid turbulence model including impurities}
%\label{}

The present flux-driven 3-d multi-species isothermal gyrofluid turbulence
model includes evolution of density profile gradients and dynamically couples
the edge pedestal region with a limiter bounded scrape-off layer (SOL).

The model is based on the local gyrofluid electromagnetic model ``GEM3''
by Scott \cite{scott03,scott05} and the SOL (limiter) model by Ribeiro \&
Scott \cite{ribeiro05,ribeiro08}, applying globally consistent geometry
\cite{scott98} with a shifted metric treatment of the coordinates
\cite{scott01}. An Arakawa-Karniadakis numerical scheme
\cite{arakawa66,karniadakis91,naulin03} is used for the computations.
The present multi-species code (``TOEFL'') has been cross-verified in the
local cold-ion limit with the tokamak edge turbulence standard benchmark case
of Falchetto et al.~\cite{falchetto08}, and with the results of finite Larmor
radius (warm ion) SOL blob simulations of Madsen~\cite{madsen11}.

In the local (delta-f) isothermal multi-species gyrofluid model \cite{scott05} the
normalised equations for the fluctuating gyrocenter densities $n_s$, including
evolution of the profile gradients, are
\begin{equation}
\partial_t n_s + [\phi_s, n_s]  =  \nabla_{||} v_{s||} + \kappa (
\phi_s +  \tau_s n_s) + S_{ns}
\label{e:den}
\end{equation}
where the index $s$ denotes the species with $s \in (e, i)$ for the main plasma
components (electrons and here main deuterium ions) plus one or more
additional ion species ($s \equiv z$).
The parallel velocities $v_{||s}$ and the vector potential $A_{||}$ evolve
according to 
\begin{equation} 
{\hat \beta} \; \partial_t A_{||} + {\hat \mu_s} ( \partial_t v_{s||} + [\phi_s,
  v_{s||}] ) =  - \nabla_{||} ( \phi_s +  \tau_s n_s )  
  + 2  {\hat \mu_s} \tau_s  \kappa ( v_{s||} ) - {\hat C} J_{||}.
\label{e:vel}
\end{equation}
The gyrocenter densities are coupled to the electrostatic potential $\phi$ by the local
gyrofluid polarisation equation
\begin{equation}
\sum_s  a_s \left[ \Gamma_{1s} n_s + \frac{1}{\tau_s} (\Gamma_{0s} -1) \phi \right]  =  0
\label{e:pol}
\end{equation}
and the velocities and current to the parallel component of the fluctuating vector
potential by Ampere's equation
\begin{equation}
\nabla_{\perp}^2 A_{||}   =  J_{||}  =  \sum_s a_s v_{s||}.
\label{e:amp}
\end{equation}
where the gyro-averaging operators in Pad\'e approximation are defined by
$\Gamma_{0s} = (1 + b_s)^{-1}$ and  $\Gamma_{1s} = (1 + (1/2) b_s)^{-1}$  with
$b_s = \tau_s \mu_s \nabla_{\perp}^2$. 

Spatial scales are normalised by the
drift scale $\rho_0 = \sqrt{T_e m_i}/(eB)$, where $T_e$ is a reference
electron temperature, $m_i$ is the ion mass, and $B$ is the magnetic field strength. 
Time scales are normalized by $c_s / L_{\perp}$, where $c_s = \sqrt{T_e/m_i}$,
and $L_{\perp}$ is the generalized profile gradient scale length. 

The parameter $a_s = Z_s n_{s0}/ n_{e0}$ describes the ratio of species
normalising background densities $n_{s0}$ to the reference density $n_{e0}$
(here usually taken at mid-pedestal value) for species with charge state
$Z_s$. The mass ratio is given by $\mu_s = m_s /(Z_s m_i)$, and the
(constant) temperature ratio by $\tau_s = T_s / (Z_s T_e)$. 

For electrons $a_e = \tau_e = -1$, $\mu_e \approx 0$, and finite Larmor radius
(FLR) effects are neglected so that $b_e \equiv 0$.
The gyro-screened potentials acting on the ions are given by $\phi_s = \Gamma_s \phi$.

Defining ${\hat \epsilon} = (qR/L_{\perp})^2$ as the squared ratio between
parallel length scale $L_{||} = q R$ (for given safety factor $q$ and torus
radius $R$) and perpendicular scale $L_{\perp}$, the main parameters 
are  ${\hat \mu}_s = \mu_s {\hat \epsilon}$, $\hat \beta = (n_{e0} T_e /
B_0^2) \hat \epsilon$, and $\hat C = 0.51 (m_e \nu_e L_{\perp} / c_{s0})
\hat \epsilon$.  
Quasi-neutrality implies $a_i = 1-a_z$. If gradients were to be fixed with
$g_s \equiv \partial_x n_{s0}$ (alternatively to the present fixed source flux
computations) then also $g_i = (1- a_z g_z)/(1- a_z)$ needs to be
satisfied. For flux-driven computations the sources have to obey
quasi-neutrality and ensure vorticity free injection. 

The nonlinear advection terms in equations (\ref{e:den}) and (\ref{e:vel}) are
expressed through Poisson brackets using the notation $[a,b] = (\partial_x
a)(\partial_y b) - (\partial_y a)(\partial_x b)$. 
Normal and geodesic components of the magnetic curvature enter the
compressional effect on vortices due to magnetic field inhomogeneity by  
%\begin{equation}
$\kappa = \kappa_y  \partial_y + \kappa_x  \partial_x$
%\end{equation}
where the curvature components in toroidal geometry are a function of the
poloidal angle $\theta$ mapped onto the parallel coordinate $z$. 
For a circular torus  $\kappa_y \equiv \kappa_0 \cos(\theta)$ and $\kappa_x
\equiv \kappa_0 \sin(\theta)$ when $\theta=0$ is defined at the outboard midplane. 

The term $S_{ns}$ on the right hand side of the density equation (\ref{e:den})
describes particle sources and sinks.  
For electrons a constant core flux driven density source is applied, which is
localized around the inside (left) radial computational boundary at $r_0$
following a narrow Gaussian profile $S_{ne} \sim s_e(r-r_0)$. 
The corresponding vorticity free source function for warm ions is FLR
corrected by $s_i = s_e - (1/2) \tau_i \nabla^2 s_e$.

The source (and sink) term $S_{nz}$ for impurity ions is here dynamically set
by ionization processes of a neutral impurity cloud, as described in the next
section. For quasi-neutral non-trace impurities the electron density has to include a
corresponding ionization source term. In the following we however will
consider only trace impurities ($a_z \ll 1$) which do not enter into
polarization or react back on the convecting main plasma turbulence. 

\medskip

For the moment we also neglect recombination 
processes. Vorticity free Dirichlet conditions are applied for all
species on the outer (right) computational boundary.
In ref.~\cite{mueller09} the necessity of using consistent energy sources has
been stressed for cases when temperature fluctuations are dynamically evolved in
addition to density fluctuations. Here we presently focus on an isothermal
model, so we do not encounter difficulties due to spurious thermal energy sources.

The present „local“ model assumes small density fluctuations on a constant
background to approximate $\ln n_s \approx n_s / n_{s0}\equiv n_s$, and can
not capture spatially localised non-trace impurity effects, like back-reaction
of impurity aggregation on the 
vortex or zonal flow fine structure.  A full-f density (``global'') model would
have to account for nonlinear (nonlocal) polarisation via  
\begin{equation}
\sum_s [q_s \Gamma_{1s} n_s + \bnabla \cdot (n_s \mu_s \bnabla) \phi ] = 0.
\end{equation}

For passively advected trace impurities ($a_z \ll 1$) we can however
consistenly solve the global gyrocenter density equation 
\begin{equation}
\partial_t n_z + [\phi_z, n_z]  =  \nabla_{||} ( n_z v_{z||} ) + n_z \kappa
(\phi_z ) +  \tau_z \kappa ( n_z ) + S_{nz} 
\label{e:nzg}
\end{equation}
and obtain the actual trace impurity particle density including nonlinear
polarisation by 
\begin{equation}
N_z = \Gamma_{1z} n_z + \frac{\mu_z}{q_z} \; \bnabla \cdot \left( n_z
\bnabla \right) \phi.  
\end{equation}
In the present simulations the local equations (\ref{e:den}), (\ref{e:vel}),
(\ref{e:pol}), (\ref{e:amp}) are used for the main plasma species, and the
global model eq.~(\ref{e:nzg}) to evolve the trace impurities. 

\newpage
\section{Dynamical source of impurity ions}
%\label{}

For simulations of trace impurity ion transport in a turbulent edge plasma, usually
either a cloud of impurity ions is initialized at a single point in time with
a given spatial distribution on a turbulent background plasma, or a constant
source of impurities is applied at each time step. The evolution of impurities
by nonlinear $E \times B$ advection is then further followed to study its
migration and inward transport, like in fluid computations of edge turbulence
of e.g. refs.~\cite{naulin05,priego05,dubuit07,futatani12}.  

Here we extend such previous approaches by applying a edge-SOL coupled model
with global profile gradient evolution, and in addition by including
a time dependent impurity ion source dynamically computed from ionization
processes depending on the local fluctuating electron density and temperature. 
A localized and (on the turbulent time scale) stationary neutral impurity
cloud acts as the background source. 

In the following we exemplarily consider electron impact ionization as the
main ionization channel. 
The temperature dependent reaction rates $R(T) = \langle \sigma v
\rangle$ of the ionization processes are (pre-)computed from fit functions of
measured or calculated ionization cross sections $\sigma(E)$.

The ionization source function $S_{nz}= \nu_{ion} n_e$ in eq.~(\ref{e:nzg})
is determined by the product of the ionization frequency $\nu_{ion}$
with the spatio-temporally fluctuating electron density $n_e({\bf x},t)$. 
The ionization frequency is determined by the neutral background gas density
$N_n$ and by the reaction rate $R(T)$ through $\nu_{ion} = N_n (r) \; R(T)$. 
The neutral density is here assumed to be constant in time and radially localized,
described by a Gaussian radial distribution $N_n (r) = N_{n0} \exp [(r-r_n)^2 /
  \Delta_n^2]$ with half width $\Delta_n$ around a position $r_n$ within the 
scrape-off-layer.

The electron impact ionization reaction rate is given by
\begin{equation}
R(T) = \langle \sigma v \rangle = \int dE f(E,T) \; \sigma(E) \; v 
= R_0 T^{-3/2} \int dE E \exp\left(-{E\over T}\right) \sigma(E).
\label{e:rate}
\end{equation}
The distribution function $f(E,T)$ is here assumed to be Maxwellian for a given
thermalized electron temperature $T$.
The prefactor $R_0$ absorbes all the constants of the Maxwell-Boltzmann
distribution and the drift scale normalization of the gyrofluid model.

The present isothermal turbulence model does not actually dynamically evolve
the electron temperature. From six-moment gyrofluid equations like in
ref.~\cite{kendl10} we can infer that the turbulent fluctuations of the
electron temperature relatively closely follow the spatio-temporal structure
of the fluctuating electron density. We can thus here approximately use a
proportionality relation $T({\bf x},t) = (T_0/n_{e0}) \; n_e({\bf x},t)$ to
dynamically compute the temperature entering in the ionization rate $R(T)$.

The ionization cross section $\sigma(E)$ in eq.~(\ref{e:rate}) may most
conveniently with sufficient accuracy be supplied in the form of fitting
functions of calculated or experimentally determined data. 

For theoretical calculations of both single atoms and also complex molecules,
the semi-classical Deutsch-M\"ark (DM) formalism
\cite{deutsch87,deutsch95,deutsch09} for electron impact ionization 
has been developed by Hans Deutsch and Tilmann M\"ark in the late 1980s
to 1990s with further following refinements and generalizations. The DM method
is based on a semi-empirical combination of the binary encounter approximation
and the Born-Bethe approximation, and has been proven of considerable
practical value.   

Here we are specifically interested in ionization cross sections of impurity
atoms and molecules relevant to fusion edge plasmas. The impurities usually
result from plasma-wall interaction processes. For carbon (or carbon fiber
composite) tiled walls or divertor plates, a multitude of different hydrocarbon
molecules and ions can form. Beryllium is intended to be used for covering the
first wall of the main chamber in ITER. Tungsten has recently received growing
interest as a potential general tokamak first wall material, and is going to
be used on strike point areas in the ITER divertor. 

Electron impact ionization cross sections of beryllium and
beryllium hydrides have recently been calculated by the DM-formalism and the
Binary-Encounter-Bethe (BEB) formalism in ref.~\cite{maihom13}.
The theoretical results have been cast into fit functions of the form
\begin{equation}
\sigma(E) = \left(a_1 \over E\right) \left[ 1-\left({I\over E} \right)
  \right]^{a_2} \left[\ln\left({E \over I}\right)+a_3+a_4\left({I\over
    E}\right)\right] 
\end{equation}
which is given units of $10^{-16}$~cm$^2$.
Here $I$ is the threshold energy and $a_i$ are the fit coefficients. For
e.g.~single ionization of gas-phase Be, the coefficients $a_1 = 70.3260$,
$a_2=1.2341$, $a_3=1.7644$ and $a_4 = -0.7628$ with $I=9.32$~eV have been
obtained as fit to the DM theory \cite{maihom13}. 

In contrast to beryllium, a comprehensive experimental data base of measured
electron impact ionization cross section data of hydrocarbon molecules and
ions has meanwhile been established. Many such fusion relevant
experimental atomic and molecular data have over the last decade been obtained
in the lab of Tilmann M\"ark, like electron impact cross sections of methane
\cite{gluch03}, acetylene \cite{feil06}, ethylene \cite{endstrasser09} 
or propene \cite{feil06ijms}.
These measured hydrocarbon cross section data have recently been re-assessed
by Huber et al. in ref.~\cite{huber11} and cast into fit functions of the
format used in the HYDKIN \cite{reiter10,hydkin} cross section data base: 
\begin{equation}
\sigma(E) = {1 \over EI} \left[ a \ln \left({E\over I}\right) + \sum_{j=1}^N
  b_j \left( 1 - {I \over E}\right)^{j} \right]
\label{e:bell}
\end{equation}
in units of $10^{-13}$~cm$^2$. For example, the fitting parameters for the
measured partial ionization cross section of ethylene for the channel $e + C_2
H_4 \rightarrow C_2 H_4^+ +2e$ have been obtained as $a = 1.5525$, $b_1 =
-1.4257$, $b_2 = 0.3340$, $b_3 =0.1928$, $b_4 = -3.8585$ and $b_5 = 2.7727$
for $I = 10.51$~eV.

%_____________
\begin{figure} % FIG 1
\includegraphics[width=7.0cm]{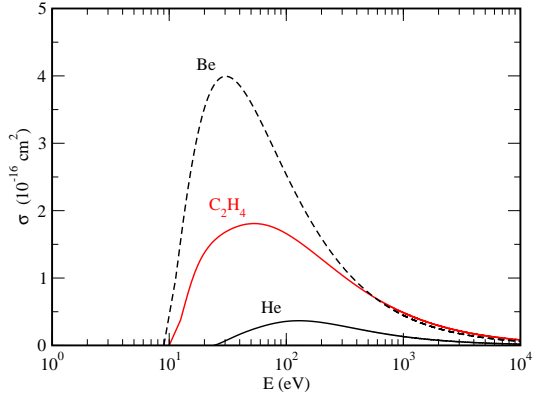}
\caption{ \sl Fit functions for the electron impact ionization cross sections of
  Be (calculated by the Deutsch-M\"ark formalism \cite{maihom13}), C$_2$H$_4$
  (fitted to  experimental data of M\"ark et al. \cite{huber11}) and He
  \cite{ralchenko08}. }  
\label{f:fig-sigma}
\end{figure}

Impurities in fusion edge plasmas may not only arise from unwanted chemical
erosion of wall materials, but can also be injected on purpose into the edge
plasma, for example as a means to enhance radiative energy losses (and thus
reduce heat fluxes) in front of the divertor by gas seeding
\cite{kallenbach14}, or for localized detection of light emissions by fast helium
gas puff imaging experiments on edge turbulence \cite{maqueda03}.

Electron impact ionization cross sections for helium are, in contrast to the previous
examples, long known with acceptable accuracy. The approximate analytic
fitting formula in the form used also above for eq.~\ref{e:bell} has been
introduced by Bell et al. in ref.~\cite{bell83}. Here we use more recent data
for electron impact ionization of helium by Ralchenko et
al.~\cite{ralchenko08}, where e.g. the parameters for single ionization of He I
from the $1^1s$ state are given as $a = 0.5857$, $b_1 = -0.4457$, $b_2 =
0.7680$, $b_3 = -2.521$, $b_4 = 3.317$ and $b_5 = 0$ with $I = 24.6$~eV. 

These fit formulas for $\sigma(E)$ are shown in fig.~\ref{f:fig-sigma} and are in the
following used to determine reaction rates $R(T)$ for the impurity ion source
terms in plasma edge turbulence simulations.

\section{Ionization and convection of Be by radially propagating blobs}
%\label{}

Cross-field transport in the scrape-off-layer of fusion plasmas is
predominantly carried by intermittent radially propagating filamentary
structures, often termed ``blobs'', which originate from a region around the
separatrix and are mostly sheared off from edge turbulence vortices and flows
in the outer closed field line region
\cite{kra01,yu03,garcia04,garcia06,garcia09,angus12,angus13,myra13}. 
 
The fully turbulent system including both the closed field line edge pedestal
region and the SOL with turbulent self-generated blob-like transport will be
considered in the next section.

Here we first focus on the ionization and further nonlinear advection of a
solitary blob structure on an initially neutral cloud within the SOL.
The present computations are fully 3-d including drift and FLR effects.

For the blob initial condition we specify a quasi-neutral Gaussian density
perturbation (with half width  6 $\rho_s$), accelerating and moving through a
radially localised cloud of impurities. 

The FLR and warm ion effect for $\tau_i=1$ have previously been found to
induce a pronounced up-down asymmetry on the radial blob propagation
\cite{madsen11}, whereas 3-d drift effect (with finite $\hat \epsilon$) have
been reported to lead to a combination of both radial blob and perpendicular
drift wave motion \cite{angus12,angus13}.

The plasma parameters for the blob computations used here are $\sqrt{\hat \epsilon} =
(qR/L_{\perp}) = 316$, $\hat \beta = 1$, $\hat C = 3.5$, $\tau_i=1$, and
$\kappa_0 = 0.05$, which reflect typical values found in tokamak edge plasmas.
As trace impurity species ($a_z = 0$) we exemplarily choose beryllium, and here for
simplicity investigate only the ionization into and dynamics of the singly
ionized charge state. In principle, multiple equations for either different
impurity species or for different charge states of one specific impurity
species could straightforwardly be implemented in the gyrofluid code. These
refinements will be considered in future work.

We assume that the beryllium ions are colder than the main ions or electrons
and set $\tau_z = 0$, and have $\mu_z = 4.5$  in relation to deuterium mass.
As numerical parameters, we use a box size of $n_x \times n_y = (256)^2$ and
$n_z = 8$ with a physical box dimension of $L_x = L_y = 128 \rho_s$ in units of the
drift scale. 

The present computations represent the first results combining both warm ion
and full 3-d effects (electromagnetic, drift effects, and toroidal
geometry) on scrape-off-layer blob motion.
A relevant result concerning the main plasma blob motion itself is, that for
the present scenario the $\tau_i$ and FLR effects slightly dominate the evolution of
the blob structure compared to the drift effects. This result is of
course strongly dependent on the specific parameters and will be discussed in
more detail elsewhere. For the moment we are more interested in the
proof-of-principle model for dynamical ionization by the blob front.

The evolution of the blob is here first followed until $t=40$ in time units of
$c_s / L_{\perp}$. For typical tokamak edge parameters this time corresponds to
around 100~$\mu$s. 

%_____________
\begin{figure} % FIG 2
\includegraphics[height=15.0cm,angle=270]{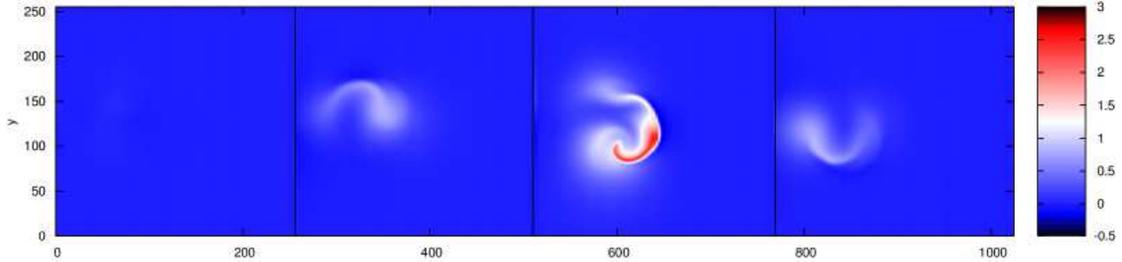}
\caption{\sl Electron density $n_e(x,y)$ filament (blob) at $t=40$ for four
  of eight parallel sections ($z=0$, $z=2$, $z=4$, $z=6$) along the field
  line. The pronouned up-down asymmetry is caused by FLR and drift effects.} 
\label{f:blob-ne}
\end{figure}

\begin{figure} % FIG 3
\includegraphics[height=6.0cm,angle=0]{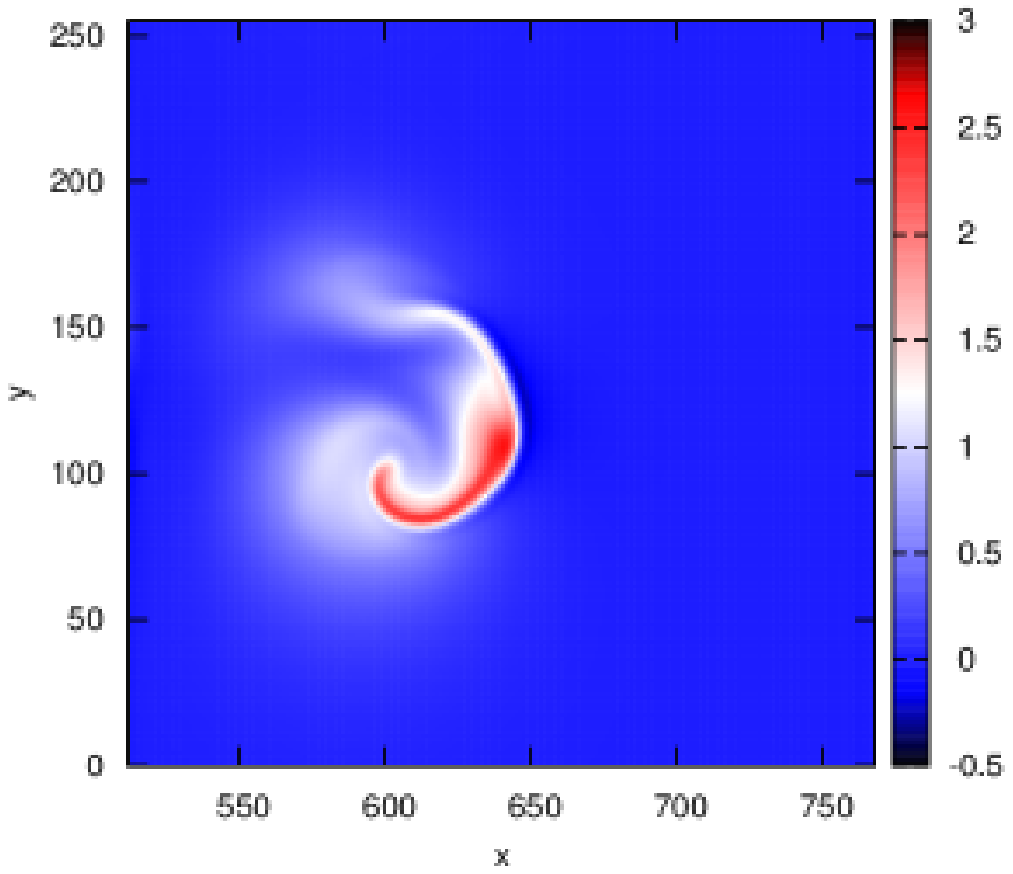}  \hspace{1.cm}
\includegraphics[height=6.0cm,angle=0]{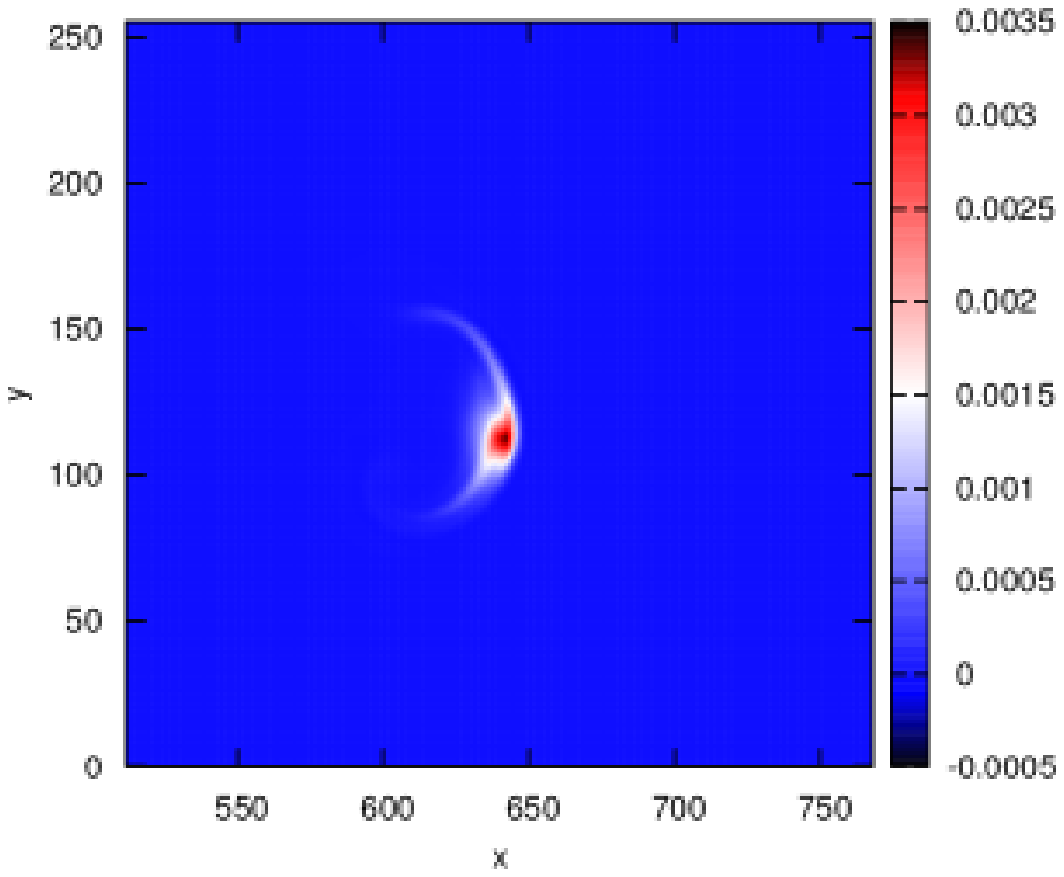} 
\caption{ \sl Left: electron density blob propagating to the right into and
  ionizing a neutral  Be cloud (not shown); right: Be$^+$ impurity ion density;
  enlarged for only the outboard  mid-plane section $z=4$ of the computational
  domain.}  
\label{f:blob-mid}
\end{figure}

In Fig.~\ref{f:blob-ne} the electron density $n_e(x,y)$ in the perpendicular
$(x,y)$ plane of the evolved blob is shown at $t=40$ for four parallel
sections ($z=0$, $z=2$, $z=4$, $z=6$) along the field line.
In Fig.~\ref{f:blob-mid} only the $z=4$ outboard midplane sections (with the
strongest interchange drive by normal curvature) for the electron density and
the Be impurity ion density are enlarged. Here the plume-like blob is
propagating to the right into a neutral Be cloud.

%_____________

\begin{figure} % FIG 4
\includegraphics[width=7.5cm]{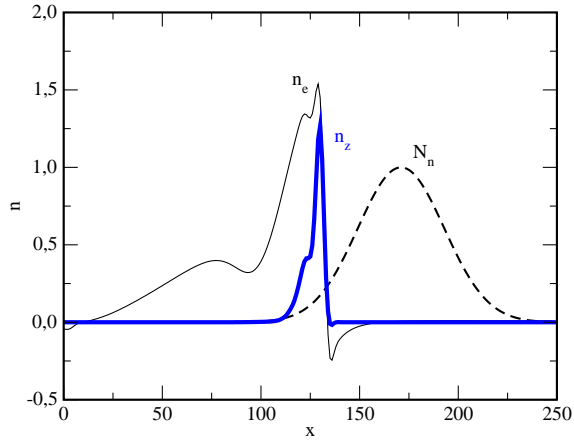} 
\caption{\sl Radial profile of the electron, impurity and neutral densities (with
  different normalizations) at $y=L_y/2$ to illustrate the situation in
  fig.\ref{f:blob-mid}.} 
\label{f:cutx}
\end{figure}

The main plasma blob shows the characteristic plume-like form (similar
to Rayleigh-B\'enard convection), with pronounced up-down asymmetries caused
by warm ion and drift effects. The beryllium ions form a more concentrated
structure, which is determined by ionization at the blob front 
propagating into the radially inhomogeneous neutral cloud, and subsequent
convection around the dipolar blob vortex.  
The densities $n(x)$ of electrons, ions and neutrals are shown (with different
specific normalizations) in fig.~\ref{f:cutx} for $y=L_y/2$ for the same
situation and time as in fig.~\ref{f:blob-mid} with different specific
normalizations.  
The impurity ion profile is mainly determined by the overlap between the
neutral cloud and the impacting blob.

%_____________

\begin{figure} % FIG 5
\includegraphics[height=6.0cm,angle=0]{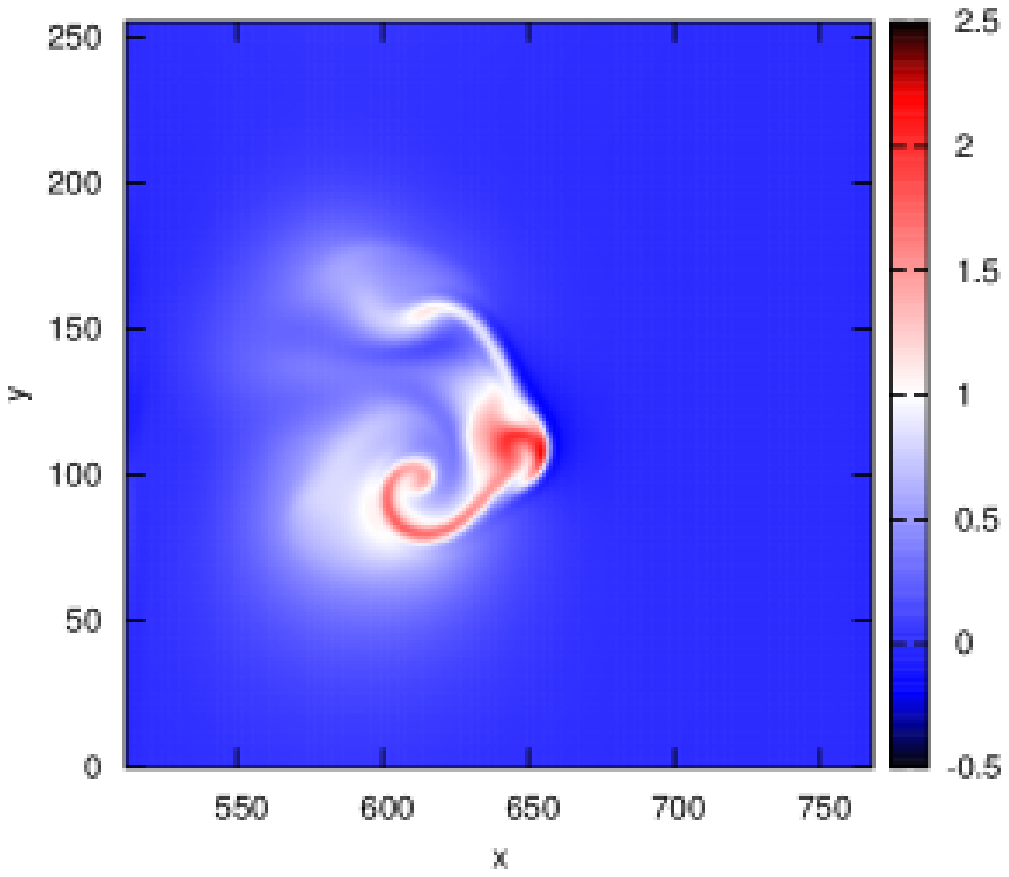} \hspace{1.cm}
\includegraphics[height=6.0cm,angle=0]{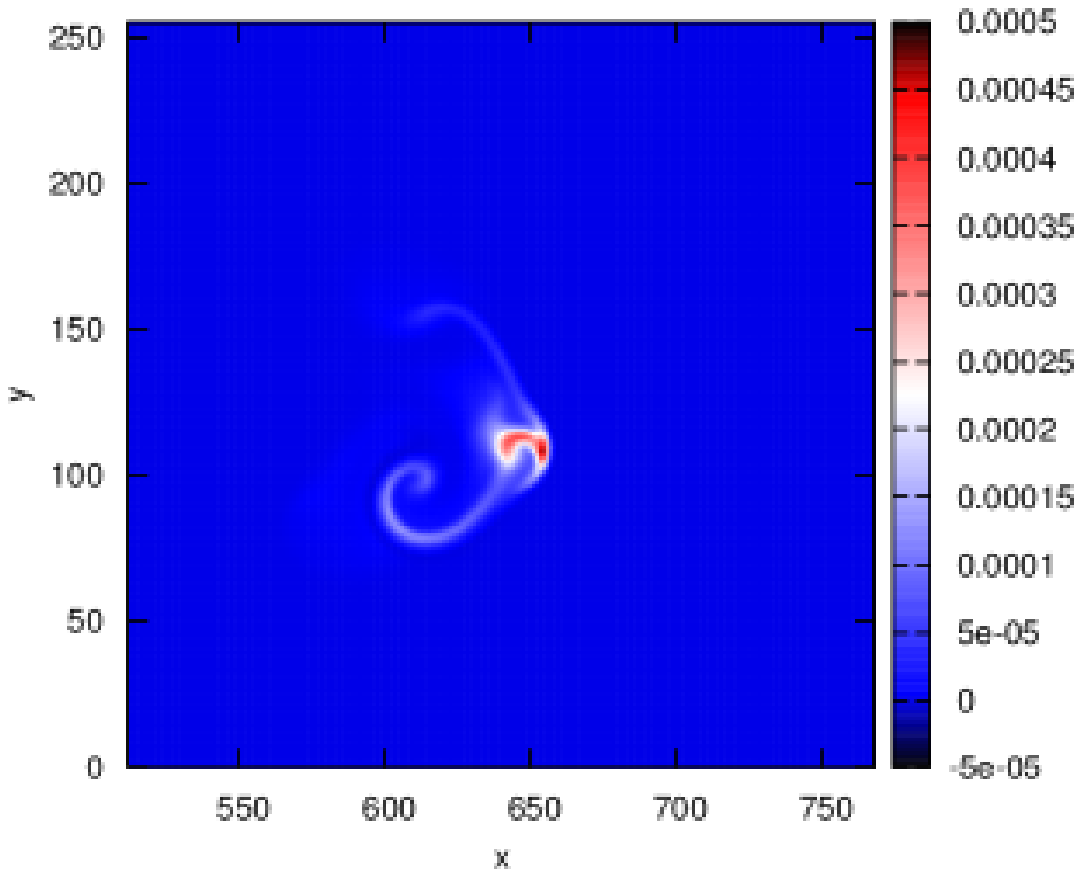} 
\caption{\sl Left: electron density blob propagating to the right across a
  constant neutral helium source. Right: product $n_z \cdot n_e$ of the
  ionized He$^+$ impurity density with the electron density. Only the outboard
  mid-plane section $z=4$ of the computational domain is shown.}  
\label{f:blob-he}
\end{figure}

This main result is similar for other dynamically ionized impuritiy species
like helium and carries significance for the interpretation of gas puff images
which follow the spatio-temporal evolution of blobs. 
As a first step towards implementing a ``virtual diagnostic'' for He puff
imaging of edge turbulence into the gyrofluid code, we plot in
fig.~\ref{f:blob-he}, for the same blob parameters as above, the product of
the electron density and helium ion density at time $t=60$. This quantity is
proportional to the recombination rate and thus is related (as a first
approximation) to the optical emission of He.  In order to separate the
eventual effects of a localized impurity source we have in this case used a
flat (constant) radial distribution of neutral helium.

%_____________

It is evident that the fast camera images delivered by imaging diagnostics in
fusion experiments may not completely capture the whole actual blob 
structure if the gas puff is inhomogeneous. For a more realistic comparison
3-d averaging effects along the line of sight and camera angle will further
have to be taken into account. 
Implementation of proper virtual diagnostics in turbulence code for
interpretation of experimental SOL transport (and, vice versa, for experimental
validation of the models and codes) is a relevant issue which has to be
considered in the appropriate detail in future work.

\section{Turbulent inward transport of hydrocarbon ions}
%\label{}

The next case is turbulence driven at the edge-core interface by a source of constant
density flux, resulting in the formation of a turbulent (L-mode like) edge
pedestal and subsequent transport of blob-like vortex structures into the SOL. 
The turbulence and profile are evolved until a saturated turbulent state is
reached (here for $t>1000$), where the average radial profiles (and energetic
turbulence quantities) remain quasi stationary. We initialize a Gaussian
neutral distribution around the right computational (wall) boundary and follow
its ionization and the further net inward convection of the resulting impurity ions. 

Parameters are $\hat \epsilon = 27000$, $\hat \beta = 1$, $\hat C = 3.5$,
$\tau_i=1$, and $\kappa_0 = 0.03$.
We now consider the single ionization of ethylene as trace impurity species
($a_z = 0$, $\tau_z = 0$, $\mu_z = 14$) as specified above.
The computational dimensions in this case are set to
$n_x \times n_y \times n_z = 128 \times 256 \times 16$ 
with $L_y = 256 \rho_s$. For the given parameters normalized to typical ASDEX
Upgrade edge conditions this corresponds to a box of approximately 5~cm width
in the radial direction, centered around the last closed flux surface, and
10~cm length in the perpendicular direction.

%_____________
\begin{figure} % FIG 6
\includegraphics[height=6.9cm]{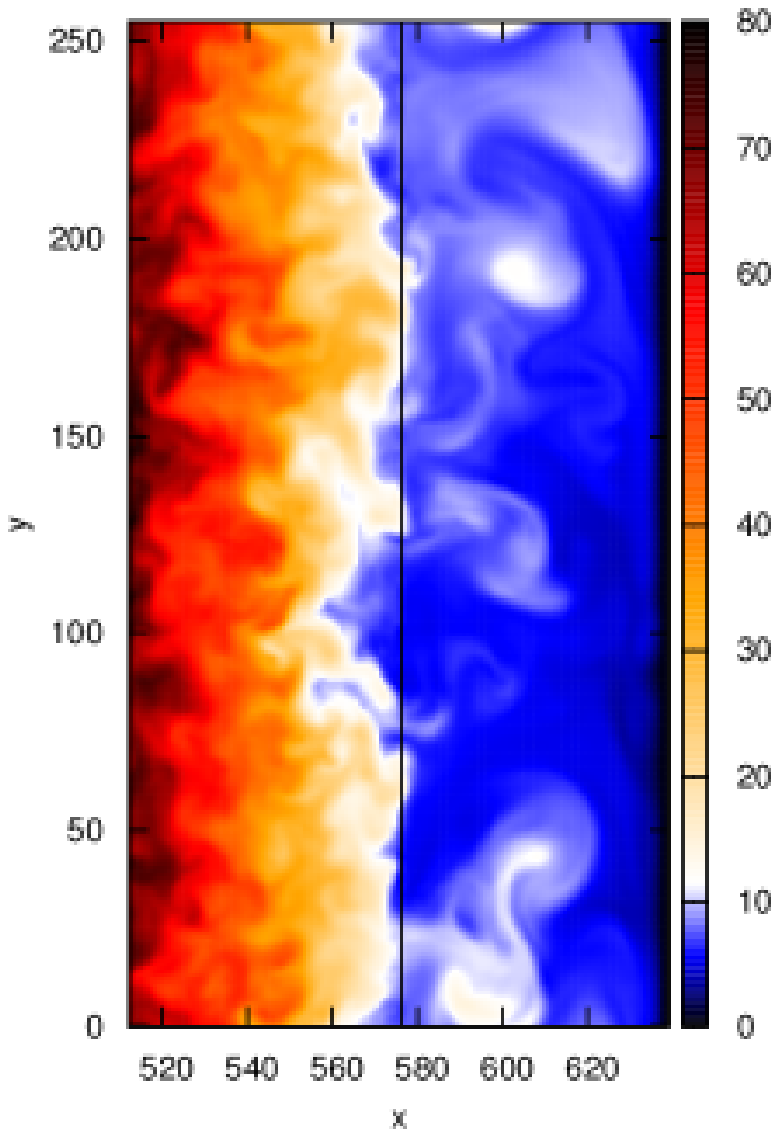}  \hspace{1.cm}
\includegraphics[height=6.9cm]{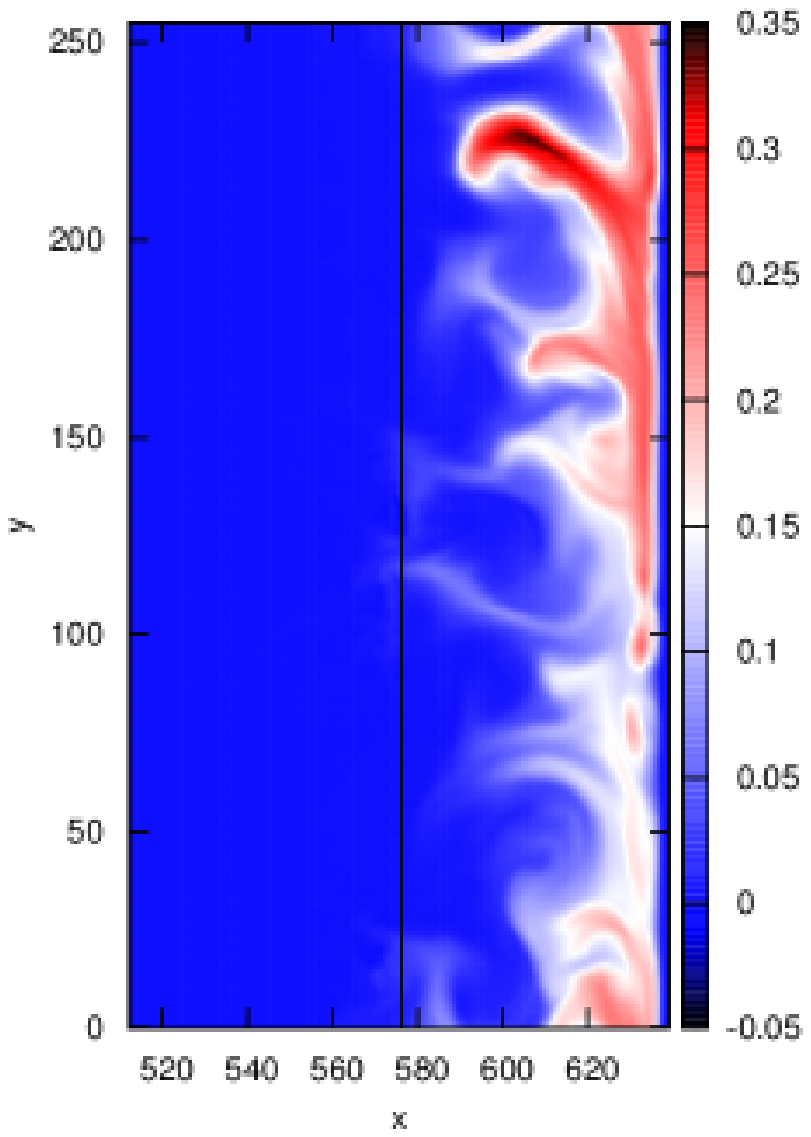} 
\caption{\sl Left: Electron density $n_e (x,y)$ in a source flux driven
  turbulence computations including profile gradient evolution from the
  pedestal top (left boundary) across the separatrix (black middle line) into
  the scrape-off layer. Right: a wall-localized neutral ethylene cloud acts as source
for electron-impact ionized $C_2H_4^+$ impurity ions density $n_z (x,y)$ which
is further transported inwards.} 
\label{f:turb-wall}
\end{figure}

Fig.~\ref{f:turb-wall} shows the electron density $n_e (x,y)$ at the left and impurity
density $n_z (x,y)$ at the right at a snapshot in time $\Delta t=40$ after
switching on the neutral source. Only the outboard midplane section $z=8$ of
16 parallel sections is here displayed. The impurities are rapidly advected
inwards from the wall with some fingers already protruding into the closed field
line region. After some time a more smoothly distributed radial impurity ion
profile will form. 

For illustration of the diffusive turbulent spreading of impurity ions from
the SOL to the closed-field line edge region we prepare an additional case,
where we inject an already pre-ionized cloud (again in the form of a
radially localized Gaussian distribution) of trace impurities into the radial middle
of the SOL domain and follow its average evolution for longer times.

\begin{figure} % FIG 7
\includegraphics[width=7.5cm]{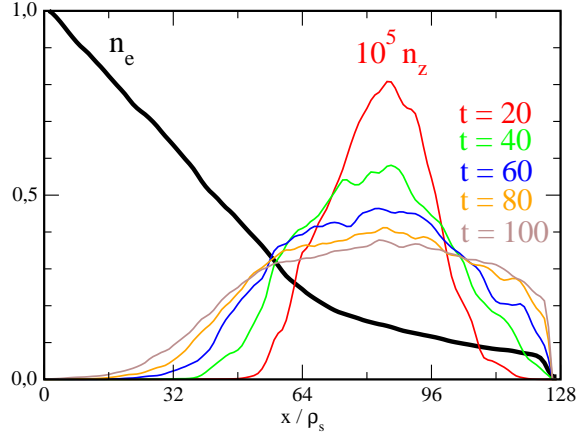} 
\caption{\sl Turbulent spreading of impurities initially localized in the SOL into
  the closed-field line main plasma region at various times after injection.}
\label{f:diffusion}
\end{figure}

Fig.~\ref{f:diffusion} shows the time evolution of $\langle n_z(x)\rangle_{y,z}$ by
its radial  spreading at five times up to $\Delta t = 100$ after injection
compared to the averaged radial profile of the electron density. The
impurities rapidly spread across the separatrix (at $x=64 \rho_s$) into the
main plasma region on a time scale of the order of cross-field blob
propagation times. 

A net pinch effect is not discernible in this figure, as the poloidal ($z$
direction) initial impurity distribution is only weakly ballooned (i.e. extended
over all $z$ planes), and the plot shows the flux-surface averaged density,
so that inward and outward curvature pinch effects (if present) mostly cancel. 
But also when the initial distribution is localized to the low-field side and
only the $y$-average of $\langle n_z(x,z_0)\rangle_y$ at this position is
taken, only a very slight inward pinch can be detected (by obtaining a
quadratic fit of the logarithm of the averaged impurity density), which is
however insignificant compared to the uncertainty present in the fit for times
before the spreading becomes influenced by the outer radial boundary.

For a further discussion of impurity pinch effects in the SOL and edge region
therefore better statistics would be necessary, by either 
strongly increasing the computational domain in the $y$ direction (and
thus getting a better $y$ average), or by studying a large number of impurity
puffs and taking additionally an ensemble average over the results.
In any case the pinch effect here appears to be small compared to the
(turbulent) diffusive spreading of the cloud. 

%_____________

\newpage
\section{Gas puff imaging of filamentary turbulent structures}
%\label{}

Now we take the Be ions out of the simulations and instead insert a helium gas puff
localized radially in the centre of the SOL region, and take a snaphot after
$\Delta t = 20$. We set up a virtual gas puff imaging (GPI) diagnostics while
taking into account dynamical ionization effects on the initially neutral gas.
The local emissivity of helium can be approximated by a power law dependence
on electron density and temperature as $I \sim n_{n0} \; n_e^{\alpha} \;
T_e^{\beta}$ with $\alpha ~\sim 0.4-0.6$ and $\beta \sim   0.6-0.8$ 
\cite{maqueda03,zweben09,russell11,shesterikov13,shao13}. 
A comparison of scrape-off-layer turbulence measured in Alcator C-Mod by probe
and (deuterium) GPI diagnostics with three-dimensional gyrofluid computations
has recently been carried out by Zweben and Scott et al.~\cite{zweben09} using
the 6-moment code GEMR. There the neutral gas density $n_{n0}$ has been
assumed to be constant in space and time over the computational area of interest. 
The GEMR computations, which also evolve temperature dynamics, further showed
that the fluctuating electron temperature is closely correlated to
the electron density \cite{zweben09}, so that the emissivity may here be well
approximated as $I \sim n_{n0} \; n_e^{\alpha + \beta}$. 

However, the neutral gas cloud density $n_{n0}(x)$ may get dynamically reduced
by local electron impact ionization from impinging turbulent blobs.
We here take this effect into account, which strongly depends on the specific
plasma parameters in the SOL.
The neutral gas density is reduced in time as
\begin{equation}
\partial_t n_n = - n_n \; n_e \; R(T) \equiv - \hat \omega_R \; n_n.
\end{equation}
The constant parameter $R_0$ in eq.~(\ref{e:rate}) is about 
$R_0 \approx 4 \cdot 10^{-9}$~cm$^3/$s when the temperature enters in units of
eV and the cross section is given in units of $10^{-16}$~cm$^2$. 
The magnitude of $R(T)$ for helium is in the order of $R_0$ for typical
near-separatrix SOL temperatures (of around 10-20 eV). 
If we assume a corresponding electron density of around 
$n_e \sim 10^{13}$~cm$^{-3}$ then the combination $\omega_R \equiv n_e \cdot R \sim
10^{4} - 10^{5}$~1/s.  
In drift normalized units $\hat \omega_R = \omega_R L_{\perp} / c_s$. For a SOL
gradient scale length of $L_{\perp} = 2$~cm and $T_e = 20$~eV in deuterium
plasma we thus have a time scale of $L_{\perp} / c_s  \sim \mu$s, and arrive
at an (order of magnitude) estimate of $\hat \omega_R \approx 10^{-2} -
10^{-1}$ for the drift normalized reduction rate.  

\begin{figure} % FIG 8
\includegraphics[height=5.85cm,angle=0]{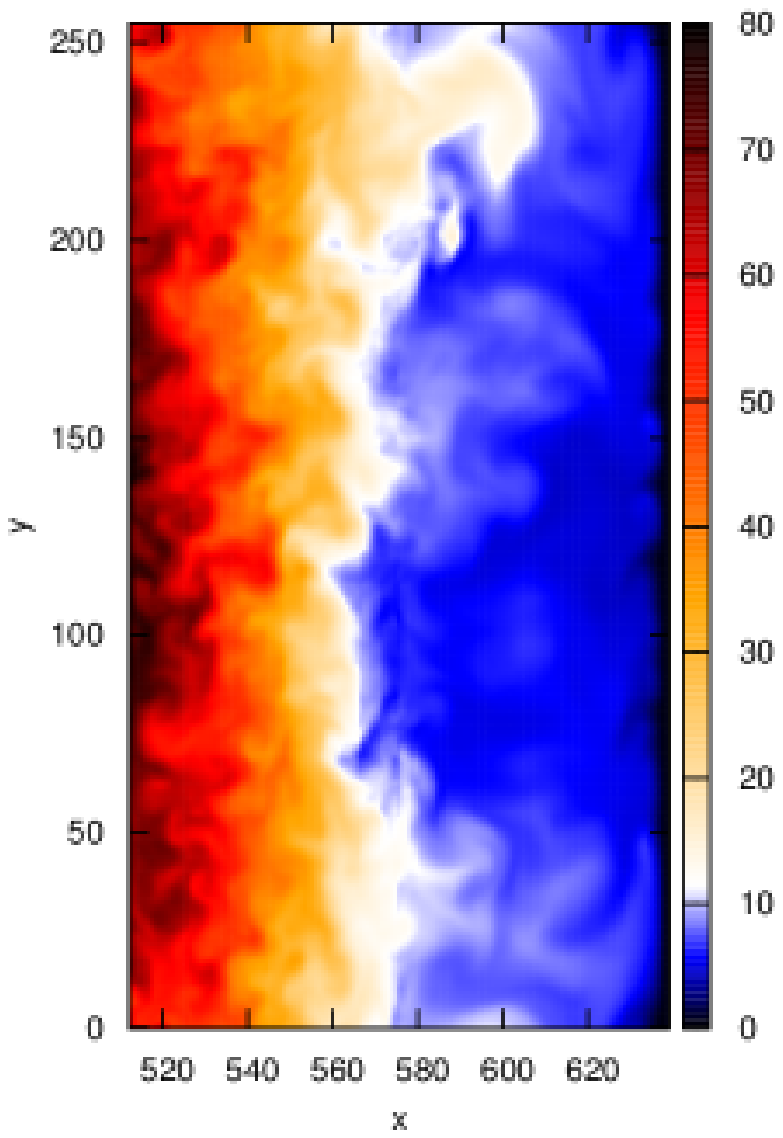} 
\includegraphics[height=5.85cm,angle=0]{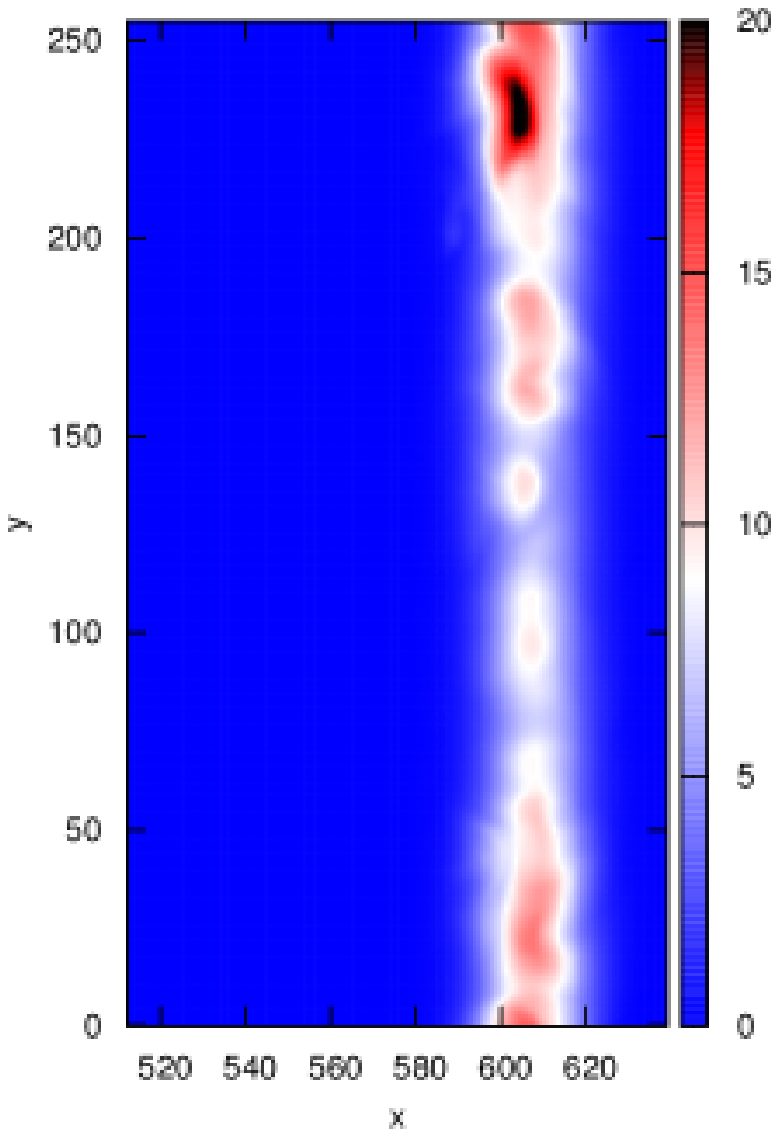} 
\includegraphics[height=5.85cm,angle=0]{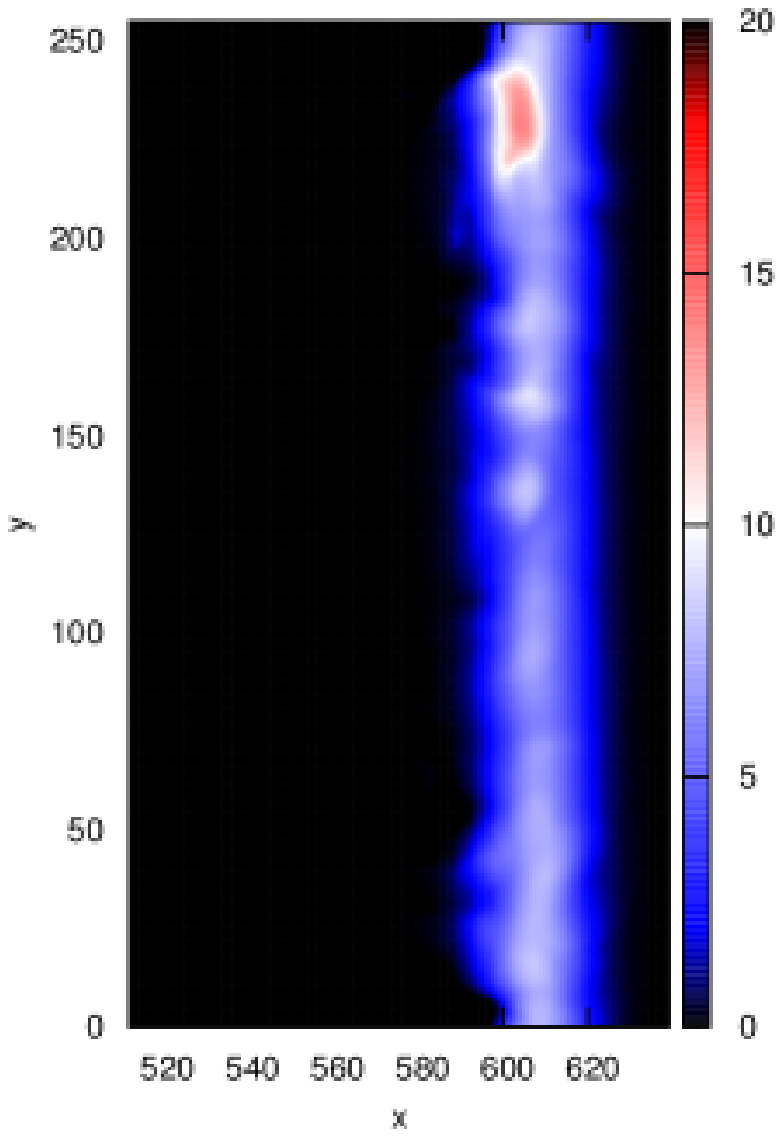} 
\includegraphics[height=5.85cm,angle=0]{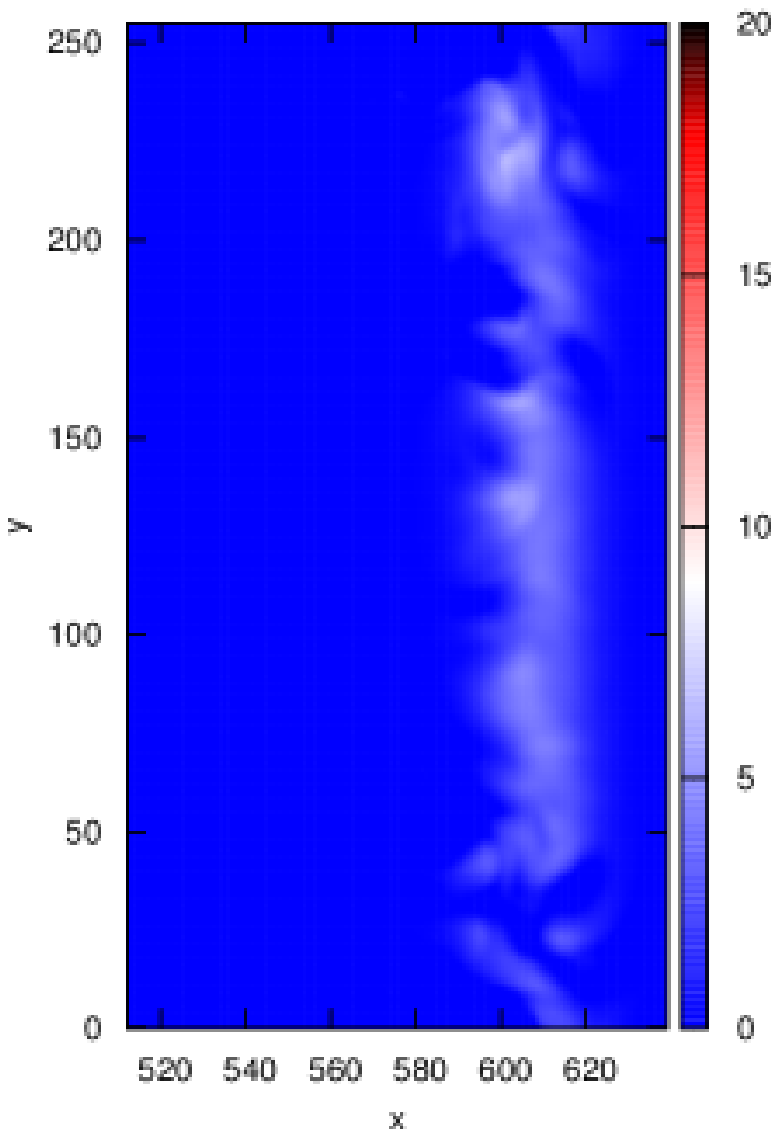} 
(a) \hspace{3.5cm} (b) \hspace{3.5cm} (c) \hspace{3.5cm} (d)
\caption{\sl A radially localized neutral helium gas cloud $n_{n0}(x)$ is puffed into the
  middle of the SOL and dynamically ionized by electron impact. (a) On top left
  a snapshot of the turbulent background electron density $n_e(x,y)$ at time
  $\Delta t = 20$ after injection is shown. The local instantaneous emission
  $I (x,y)\sim n_{n0} n_e^{\alpha+\beta}$, simulating  a virtual GPI
  diagnostic, is shown for three values of the neutral depletion rate: (b) $\omega_R = 0$,
  (c) 0.2 and (d) and 0.5.}  
\label{f:turb-puff}
\end{figure}

In fig.~\ref{f:turb-puff} we plot the electron density (top left) and the
spatial distribution of the simulated GPI emission intensity for various
values of the neutral (helium) reduction rate: $\hat \omega_R = 0$ (top right), $\hat
\omega_R = 0.2$ (bottom left), $\hat \omega_R = 0.5$ (bottom right). The
results show that not only the intensity but also the spatial structure of the
emission strongly depends on the plasma parameters: for low electron densities
the impact ionization rate is small, and over typical turbulent or blob time
scales no significant effect is visible. The image for $\hat \omega_R = 0.01$
is indistinguisable from the one for $\hat \omega_R = 0$.
For larger SOL densities and thus larger $\hat \omega_R > 0.1$ the dynamical
ionization may however distort the obtained GPI picture.

%_____________

A common caveat concerning both the GEMR code and our code is the use of local
``delta-f'' models, which is strictly only applicable radially locally in thin
layers of plasma. In a global sense the ``delta-f'' condition of small
fluctuation amplitudes compared to the background is never really fulfilled
for typical SOL conditions. 
More realistic 3-d gyrofluid code comparisons with SOL measurements in future
have to be based on ``full-f'' global edge-SOL models, which are
presently still under development \cite{madsen13}.

\newpage
\section{Conclusions}
%\label{}

We have shown how measured and calculated electron impact ionization cross
section data can be included in multi-species gyrofluid modelling of
turbulence in the edge of magnetized fusion plasmas.
Here exemplarily only one trace impurity species in addition to the main plasma
species has been implemented. In principle any number of further impurities
or charge states can be straightforwardly added.
A number of restrictions of the model remain to be resolved before any detailed
realistic comparison between the gyrofluid code results and actual
measurements in the SOL of tokamaks may be undertaken. The most important
generalization to the present models will be development of 3-d 6-moment
multi-species ``full-f'' codes including fully global evolution of the
turbulence and profiles. In such a model also the back reaction of (non-trace)
impurities on local turbulence and flow structure could be consistently included. 
Further care will have to be devoted to the implementation of realistic flux
surface geometry including a consistent treatment of the X-point and separatrix.

\section*{Acknowledgements}
%% \label{}
This work is dedicated to Tilmann M\"ark on the occasion of his 70th Birthday
Honor Issue of the International Journal of Mass Spectrometry.
The work was partly supported by the Austrian Science Fund (FWF) Y398, and
by the European Commission under the Contract of Association between EURATOM
and \"OAW, carried out within the framework of the European Fusion Development
Agreement (EFDA). The views and opinions expressed herein do not necessarily
reflect those of the European Commission.


\newpage

\begin{thebibliography}{00}
%\section*{References}

\bibitem{janeschitz01}
G. Janeschitz et al.,
%Plasma-wall interaction issues in ITER.
\newblock Journal of Nuclear Materials {\bf 290-293}, 1 (2001).

\bibitem{federici03}
G. Federici, P. Andrew, P. Barabaschi, et al.,
%Key ITER plasma edge and plasma-material interaction issues.
\newblock Journal of Nuclear Materials {\bf 313-316}, 11 (2003).

\bibitem{roth09}
J. Roth, E. Tsitrone, A. Loarte, et al.
\newblock Journal of Nuclear Materials {\bf 390-391}, 1 (2009).

\bibitem{clark05}
R.E.H. Clark, D.H. Reiter (eds.): {\sl Nuclear fusion research - understanding
plasma-surface interactions.} Springer, Berlin 2005.

\bibitem{tang78}
W.M. Tang, 
\newblock Nucl. Fusion {\bf 18}, 1089 (1978).

\bibitem{hugill93}
J. Hugill, 
\newblock Nucl. Fusion {\bf 23}, 331 (1983).

\bibitem{liewer85}
P.C. Liewer, 
\newblock Nucl. Fusion {\bf 25}, 543 (1985).

\bibitem{wootton90}
A.J. Wootton, B. A. Carreras, H. Matsumoto, et al.,
%K. McGuire, W. A. Peebles, C. P. Ritz, P. W. Terry, and S. J. Zweben, 
\newblock Phys. Fluids B {\bf 2}, 2879 (1990).

\bibitem{dimits00}
A.M. Dimits, G. Bateman, M. A. Beer, et al.,
%B. I. Cohen, W. Dorland, G. W. Hammett, C. Kim, J. E. Kinsey,
%M. Kotschenreuther, A. H. Kritz et al., 
\newblock Phys. Plasmas {\bf 7}, 969 (2000).

\bibitem{scott03} 
B.D. Scott, 
\newblock Plasma Phys. Contr. Fusion {\bf 45} A385 (2003).
 
\bibitem{garbet04}
X. Garbet, P. Mantica, C. Angioni, et al.
%E. Asp, Y. Baranov, C. Bourdelle, R. Budny, F. Crisanti, G. Cordey,
%L. Garzotti et al.,  
\newblock Plasma Phys. Controlled Fusion {\bf 46}, B557 (2004).

\bibitem{tynan09}
G.R. Tynan, A. Fujisawa, and G. McKee,
\newblock Plasma Phys. Control. Fuson {\bf 51}, 113001 (2009).
 
\bibitem{diamond05}
P H. Diamond, S.I. Itoh, K. Itoh, and T.S. Hahm, 
\newblock Plasma Phys. Controlled Fusion {\bf 47}, R35 (2005).

\bibitem{kendl10}
A. Kendl, B. Scott and T.T. Ribeiro, 
%Nonlinear gyrofluid computation of edge localized ideal ballooning modes,
\newblock Phys. Plasmas {\bf 17}, 072302 (2010).

\bibitem{scott05} 
B.D. Scott, 
\newblock Phys. Plasmas {\bf12} 102307 (2005).

\bibitem{ribeiro05} 
T.T. Ribeiro and B.D. Scott, 
\newblock Plasma Phys. Control. Fusion {\bf47} 1657 (2005).
    
\bibitem{ribeiro08} 
T.T. Ribeiro and B.D. Scott, 
\newblock Plasma Phys. Control. Fusion {\bf50} 055007 (2008)
   
\bibitem{scott98} 
B.D. Scott, 
\newblock Phys. Plasmas {\bf 5} 2334 (1998).
    
\bibitem{scott01} 
B.D. Scott, 
\newblock Phys. Plasmas {\bf8} 447 (2001).

\bibitem{karniadakis91}
G.E. Karniadakis, M. Israeli, S.A. Orszag,
\newblock  J. Comput. Phys. {\bf 97}, 414 (1991).

\bibitem{arakawa66}
A. Arakawa,
\newblock J. Comput. Phys. {\bf 1}, 119 (1966).

\bibitem{naulin03}
V. Naulin, A. Nielsen,
\newblock SIAM J. Sci Comput. {\bf 25}, 104 (2003).

\bibitem{falchetto08}
G.L. Falchetto, B.D. Scott, P. Angelino, et al.,
%A. Bottino, T. Dannert, V. Grandgirard, S. Janhunen, F. Jenko, S. Jolliet,
%A. Kendl, B.F. McMillan, V. Naulin, A.H. Nielsen, M. Ottaviani, A.G. Peeters,
%M.J. Pueschel, D. Reiser, T.T. Ribeiro and M. Romanelli, 
\newblock Plasma Phys. Control. Fusion 50, 124015, 12pp (2008).

\bibitem{madsen11}
J. Madsen, O.E. Garcia, J.S. Larsen, et al.,
%The influence of finite Larmor radius effects on the radial interchange
%motion of plasma filaments. 
\newblock Phys. Plasmas {\bf 18}, 112504 (2011).

\bibitem{naulin05}
V. Naulin, 
\newblock Phys. Rev. E {\bf 71}, 015402 (2005).


\bibitem{priego05}
M. Priego, O.E. Garcia, V. Naulin, J.J.Rasmussen,
%Anomlaous diffusion, clustering and pinch of impurities in plasma edge
%turbulence 
Phys. Plasmas {\bf 12}, 062312 (2005).

\bibitem{dubuit07}
N. Dubuit, X. Garbet, T. Parisot et al,
\newblock Phys. Plasmas {\bf 14}, 042301 (2007).

\bibitem{futatani12}
S. Futatani, D.del-Castillo-Negrete, X. Garbet et al.,
\newblock Phys. Rev. Lett. {\bf 109}, 185005 (2012).

\bibitem{deutsch87}
H. Deutsch and T.D. M\"ark,
\newblock Int. J. Mass Spectrom. Ion Proc. {\bf 79}, R1 (1987).

\bibitem{deutsch95}
H. Deutsch, K. Becker, T.D. M\"ark,
\newblock Contrib. Plasma Phys. {\bf 35}, 421 (1995)

\bibitem{deutsch09}
H. Deutsch, K. Becker, M. Probst, T.D. M\"ark,
\newblock Advances in Atomic, Molecular and Optical Physics {\bf 57}, 87 (2009).

\bibitem{becker88}
K.Becker, F.Biasioli, G.Denifl, et al., 
%H.Deutsch, T.Fiegele, V.Grill, T.D.Märk, C.Mair, S.Matt, D.Muigg, P.Scheier, M.Sonderegger, A.Stamatovic, R.Wörgötter.
%Electron impact ionization and surface induced reactions of fusion plasma
%edge constituents,
\newblock NIST Special Publications, {\bf 926}, 205-209 (1988).


\bibitem{Maerk92}
T.D. M\"ark,
%Ionization by electron impact 
\newblock Plasma Phys. Control. Fusion {\bf 34}, 2083 (1992).


\bibitem{mueller09}
S.H. M\"uller, C. Holland, G.R. Tynan, J.H. Hu and V. Naulin, 
%Source formulation for electron-impact ionization for fluid plasma simulations
\newblock Plasma Phys. Control. Fusion {\bf 51}, 105014 (2009)


\bibitem{maihom13}
T. Maihom I. Sukuba, R. Janev, et al.,
\newblock Eur. Phys. J. D. {\bf 67}, 2 (2013).

\bibitem{gluch03}
K. Gluch, P. Scheier, W. Schustereder, et al.,
\newblock Int. J. Mass Spectrom. {\bf 228}, 307 (2003).

\bibitem{feil06}
S. Feil, K. Gluch, A. Bacher, et al.,
\newblock J. Chem. Phys. {\bf 124}, 214307 (2006).

\bibitem{endstrasser09}
N. Endstrasser, F. Zappa, A. Mauracher, et al.,
\newblock Int. J. Mass Spectrom. {\bf 280}, 65 (2009).

\bibitem{feil06ijms}
S. Feil, A. Bacher, K. Gluch, et al.,
\newblock Int. J. Mass Spectrom. {\bf 253}, 122 (2006).

\bibitem{huber11}
S.E. Huber, J. Seebacher, A. Kendl and D. Reiter,
\newblock Contrib. Plasma Phys. {\bf 51}, 931 (2011).

\bibitem{reiter10}
D. Reiter and R.K. Janev,
\newblock Contrib. Plasma Phys. {\bf 50} 986 (2010).

\bibitem{hydkin}
D. Reiter and B. K\"uppers,
\newblock http://www.hydkin.de (2010).

\bibitem{kallenbach14}
A. Kallenbach, M. Bernert, R. Dux, et al.
\newblock {\sl Impurity seeding for tokamak power exhaust: from present devices
  via ITER to DEMO}, Plasma Phys. Contr. Fusion (2014), in print.

\bibitem{maqueda03}
R.J. Maqueda, G.A. Wurden, D.P. Stotler et al.,
\newblock Rev. Sci. Instrum. {\bf 74}, 2020 (2003).

\bibitem{bell83}
K.L. Bell, H.B. Gilbody, J.G. Hughes, et al.,
\newblock L. Phys. Chem. Ref. Data {\bf 12}, 891 (1983).

\bibitem{ralchenko08}
Y. Ralchenko, R. Janev, T. Kato et al.,
\newblock Atomic and Nuclear Data tables {\bf 94}, 603 (2008).


\bibitem{kra01}
S.I. Krasheninnikov,
\newblock Phys. Lett. A {\bf 283}, 368 (2001).

\bibitem{yu03}
G.Q. Yu and S.I. Krasheninnikov,
\newblock Phys. Plasmas {\bf 10}, 4413 (2003).

\bibitem{garcia04}
O.E. Garcia, V. Naulin, A.H. Nielsen and J.J. Rasmussen,
\newblock Phys. Rev. Lett. {\bf 92}, 165003 (2004).

%\bibitem{garcia05}
%O.E. Garcia, V. Naulin, A.H. Nielsen and J.J. Rasmussen,
%\newblock  Phys. Plasmas {\bf 12}, 062309 (2005)

\bibitem{garcia06}
O.E. Garcia, N.H. Bian and W. Fundamenski,
\newblock Phys. Plasmas {\bf 13}, 082309 (2006)

\bibitem{garcia09}
O.E. Garcia,
\newblock Plasma and Fusion Research {\bf 4}, 019 (2009).

\bibitem{angus12}
J.R. Angus, M.V. Umansky, and S.I. Krasheninnikov,
\newblock Phys. Rev. Lett. {\bf 108}, 215002 (2012).

\bibitem{angus13}
J.R. Angus, M.V. Umansky, and S.I. Krasheninnikov,
\newblock J. Nucl. Materials {\bf 438}, S572 (2013).

\bibitem{myra13}
J.R. Myra, W.M. Davis, D.A. D'Ippolito et al.,
\newblock Nucl. Fusion {\bf 53}, 073013 (2013).

\bibitem{zweben09}
S.J. Zweben, B.D. Scott, J.L. Terry et al., 
\newblock Phys. Plasmas {\bf 16}, 082505 (2009).

\bibitem{russell11}
D.A. Russell, J.R. Myra, D.A. D'Ippolito, et al.,
\newblock Phys. Plasmas {\bf 18}, 022306 (2011).

\bibitem{shesterikov13}
I. Shesterikov, Y. Xu, M. Berte, et al.,
\newblock Review of Scientific Instruments {\bf 84}, 053501 (2013).

\bibitem{shao13}
L.M. Shao, G.S. Xu, S.C. Liu, et al.
\newblock Plasma Phys. Control. Fusion {\bf 55}, 105006 (2013)

\bibitem{madsen13}
J. Madsen,
\newblock Phys. Plasmas {\bf 20}, 072301 (2013).


\end{thebibliography}
\end{document}